\documentclass[11pt,twoside]{article}

\usepackage{asp2004}
\usepackage{graphicx}

\begin{document}
\title{Stellar Abundances: Recent and Foreseeable Trends}   
\author{Carlos Allende Prieto}   
\affil{University of Texas, Austin, TX 787512, USA}    

\begin{abstract}
The determination of chemical abundances from stellar spectra 
is considered a mature field of astrophysics. Digital spectra of stars 
are recorded and processed with standard techniques, much like samples 
in the biological sciences. Nevertheless, uncertainties typically exceed 20\%, 
and are dominated by systematic errors. The first part of this paper  
addresses what is being done to reduce measurement errors; and what is not 
being done, but should be. The second part focuses in some of the 
most exciting applications of stellar spectroscopy in the arenas of galactic 
structure and evolution, the origin of the chemical elements, and cosmology.
\end{abstract}

\section{Introduction}

Stellar absorption lines, first discovered in the solar spectrum circa 1804, 
can be used to quantify the proportions of different chemical 
elements in stars. The strength of the lines depends on the 
abundance of absorbers, but also on the environment where the absorption 
takes place: the stellar atmosphere. The determination of chemical 
abundances requires an accurate knowledge of the physical conditions 
(e.g. temperature, density, radiation field) in the outer layers of the star.
In other words, inferring chemical abundances from spectra is just a part
of the problem of the physics of stellar atmospheres, and cannot be
decoupled from it.

After Eddingon, Milne, and others established the theoretical basis, 
the field of quantitative stellar 
spectroscopy  has evolved over time to become almost an 
industry. Stellar abundances have shed light on the origin of the
elements, the early universe,  what the interior of 
the stars look like, or how galaxies form and evolve. 
As modeling is refined, so
are the quality and reach of the observations, making possible to use
stellar abundances to tackle new problems. 

Nowadays,  a small army of some $10^3$ researchers 
are {\it professionally} dedicated to measure chemical abundances 
from stellar spectra in the entire world. 
Practitioners favor B-type stars on the warm
side, and F- and G-type stars on the cool side.  B-type stars
can be observed at large distances, while the cooler F- and G-types
live long, allowing us to study the past.  Cooler stars
have far more complex spectra, shaped by molecular absorption and dust. 

In this paper, with no ambition of formally reviewing the field, 
I simply highlight recent accomplishments and tendencies, both
in methods (Section \ref{howto}) and applications (\S \ref{whatfor}). 
The closing section is devoted to a personal (and surely biased) 
reflection on where the field is going, and where one wishes it would go.

\section{Stellar abundances. How to?}
\label{howto}

Telescopes and spectrographs and used to obtain stellar spectra. 
The visible 
is the window of choice for abundance measurements, 
because of the high transparency of the Earth's atmosphere, 
limited line crowding, 
and simple continuum opacities (basically H and H$^{-}$).  
The result of an observation is an account of the detected
 stellar photons properly ordered by energy. The next step is removing 
instrumental effects that distort the observed photon distribution. 

The final stage consists in translating observed line strengths into 
chemical abundances. This step involves 
many simplifications and assumptions that allow us to
build an approximate description of the physical conditions 
 in the atmosphere of the star, given a small number of  
parameters (energy flux, surface gravity, and 
metal abundances\footnote{Hereafter 
quantified in a logarithmic scale relative
to hydrogen and solar values:

[X/H] $\displaystyle = \log  \frac{{\rm N}_{\rm X}}{{\rm N}_{\rm H}}  -
  	 \log  \left(\frac{{\rm N}_{\rm X}}{{\rm N}_{\rm H}}\right)_{\odot}$,
  	 where N represents number density.}) 
to be inferred from observations. 
Libraries of atmospheric models are computed by experts, and then 
widely disseminated (or not). The fact that one of the model parameters, 
the metal abundance, is what we seek, makes this a problem that 
needs to be treated iteratively.
In the standard procedure, the equation of radiative 
transfer is solved to calculate the spectra predicted by the models,
which are then compared to an observation 
to constrain the atmospheric parameters, and estimate the metal abundances.
This information is then used to select a new model that will be
used to refine the abundances, going on until convergence is achieved.

Two critical parts of this process are the model atmospheres 
and the calculation of the emergent spectrum from the models. 
Obtaining the right observations and performing a reliable analysis 
are other key elements involved.

\subsection{Model atmospheres}
\label{models}

The model atmospheres commonly used today are based on the concepts
of radiative and hydrostatic equilibrium. These models are 
one-dimensional: all the relevant thermodynamical quantities depend only 
 on height (plane-parallel geometry) or radius (spherical models). 
Models for late-type stars also assume local thermodynamical equilibrium (LTE), 
i.e. the source function 
is equal to the Planck law and therefore only depends on 
the local temperature.

\begin{figure}[ht!]
\centering
\includegraphics[width=9cm,angle=0]{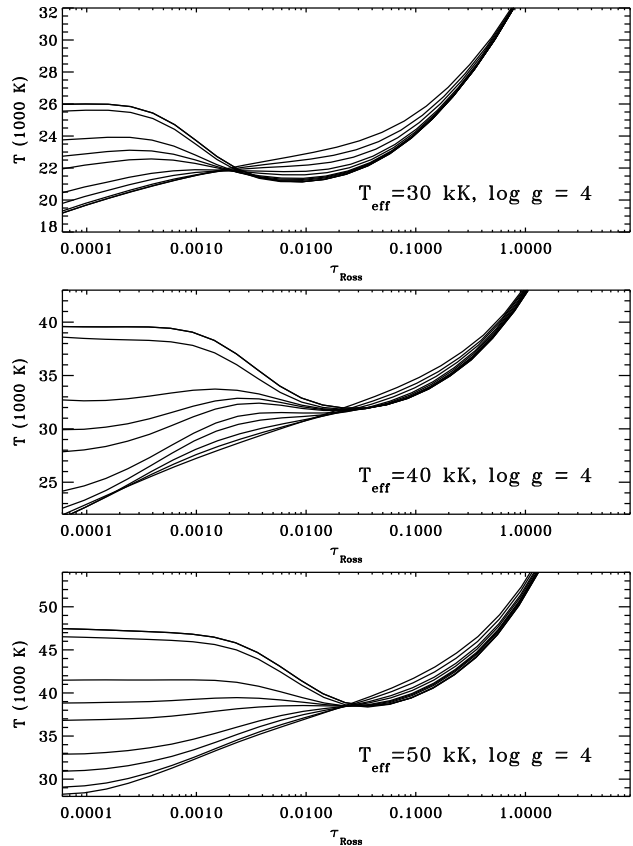}
\protect\caption[ ]{
Temperature structure as a function of the Rosseland optical depth 
from the O-star grid of Lanz \& Hubeny (2003). The different curves for 
any given $T_{\rm eff}$ correspond to metallicities from $2\times$ solar
to zero. Departures from LTE are responsible for warming the outer 
layers of the more metal-poor models, where line cooling is not effective.
}
\label{f1}
\end{figure}

Spectral lines are most useful for deriving chemical abundances, but
line absorption also affects the energy balance in the outer layers
of a star. Iron is by far the element 
that produces most  of the observed lines. Metal line 
crowding in the blue and UV spectra of late-type stars 
contributes significantly to the total opacity, inducing 
extra warming in
deep atmospheric layers and cooling in the outer regions. 
Line blanketing becomes 
milder for hot stars, but even then it is still quite significant. 
Hot star models that account
 for both departures from LTE and line blanketing (see Fig. \ref{f1})
have only become
available  recently (Hubeny \& Lanz 1995,  Lanz \& Hubeny 2003, Rauch 2003,
Repolust, Puls, J. \& Herrero 2004).
For the most massive stars, especially when dealing with UV spectra,
winds need to be considered (Pauldrach et al. 2001, Hillier 2003).

\begin{figure}[ht!]
\centering
\begin{tabular}{ccc}
&
\includegraphics[width=8cm,angle=0]{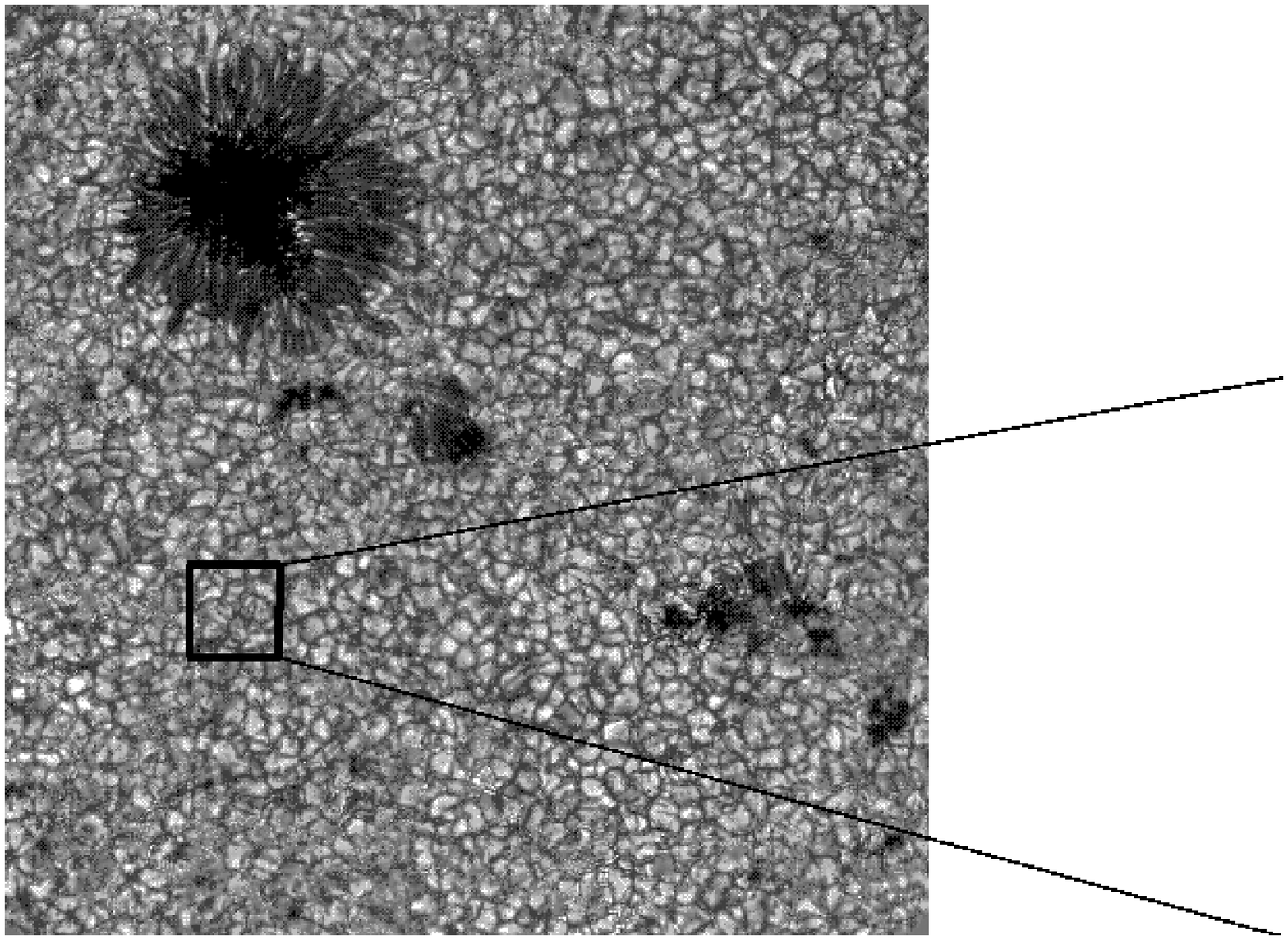} &
\includegraphics[width=5.cm,angle=0]{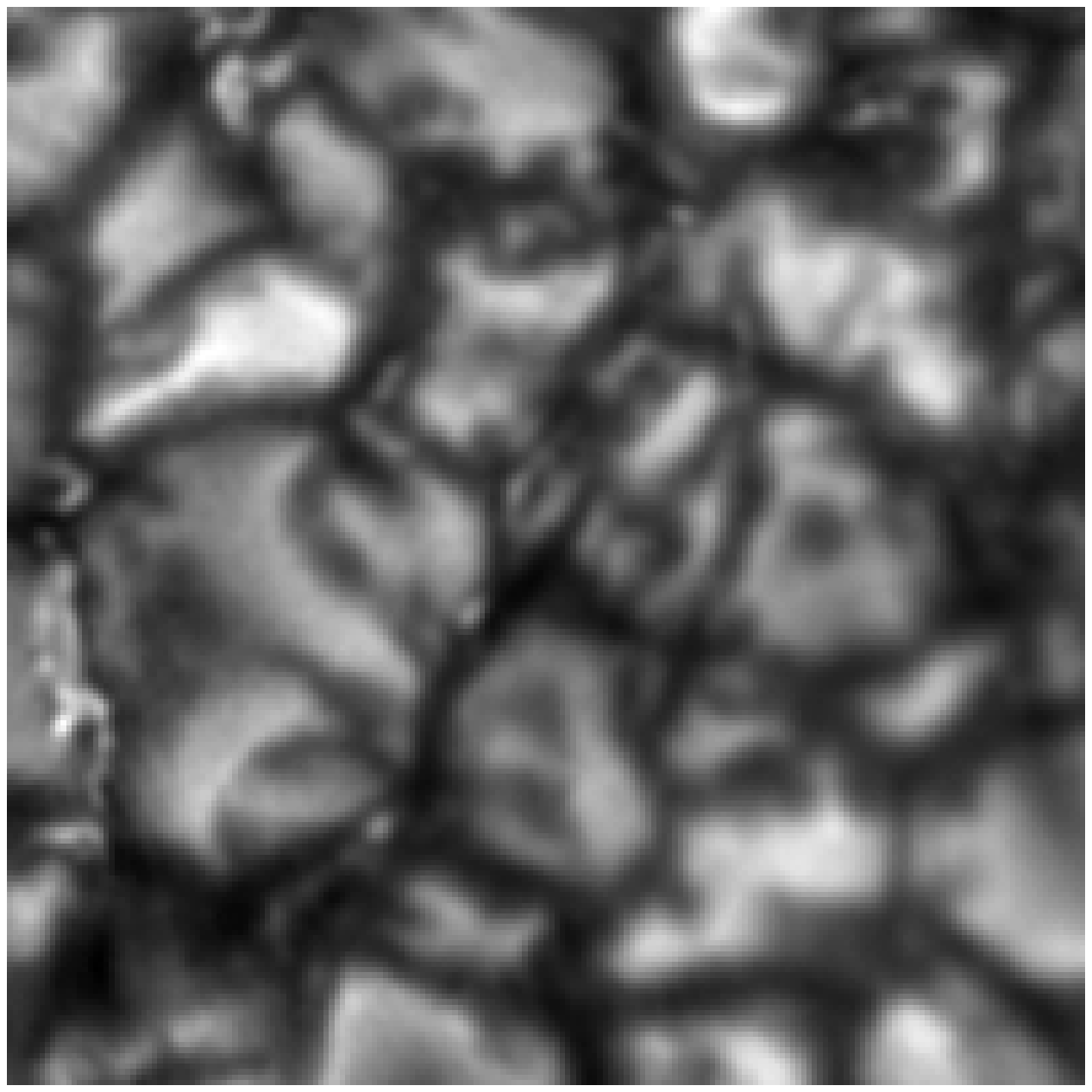}
\end{tabular}
\protect\caption[ ]{
Image of the solar surface obtained with the Swedish 1-m Solar Telescope 
at La Palma. The continuum image (436.4 nm) shows a group
of sunspots on August 22, 2003. Observations  by
O. Engvold, J. E. Wiik, and L. R. van der Voort (Inst. of Theoretical 
Astrophysics of Oslo University). 
}
\label{f2}
\end{figure}

Models for late-type (F, G, and K-type) stars have been  
always based on the assumption that LTE holds. As stated above, 
line blanketing needs to be included.
Unlike in more massive stars, convection develops in the envelopes of these stars 
to some degree. This is accounted for in the energy balance by a simplified
treatment termed as mixing-length theory (B\"ohm-Vitense 1958). 
However, the impact of convection
on the velocity fields at the surface of the star is usually ignored, i.e.,
hydrostatic equilibrium is still assumed. That this is a bad approximation 
is nowhere more evident than in a  high-resolution 
image of the solar surface. 
Fig. \ref{f2} shows an area of the solar disk including a group of sunspots. 
The concentration of magnetic field
in sunspots inhibits convection, which is responsible for the 
granular pattern apparent everywhere else, as evident 
in the section expanded in the right-hand panel of Fig. \ref{f2}.
The white areas correspond to bright granules: hot gas upflows. 
The dark lanes are the intergranules: cooler sinking gas. The temperature
contrast in the solar photosphere ranges from several hundred to more 
than a thousand degrees, 
and the velocities reach a few kilometers per second.  Typical granules 
have angular sizes of about 1  arcsecond, or a couple thousand kilometers 
on the solar surface, and evolve on time scales of the order of 10--15 minutes.

\begin{figure}[ht!]
\centering
\begin{tabular}{ccc}
&
\includegraphics[width=3.6cm,angle=0]{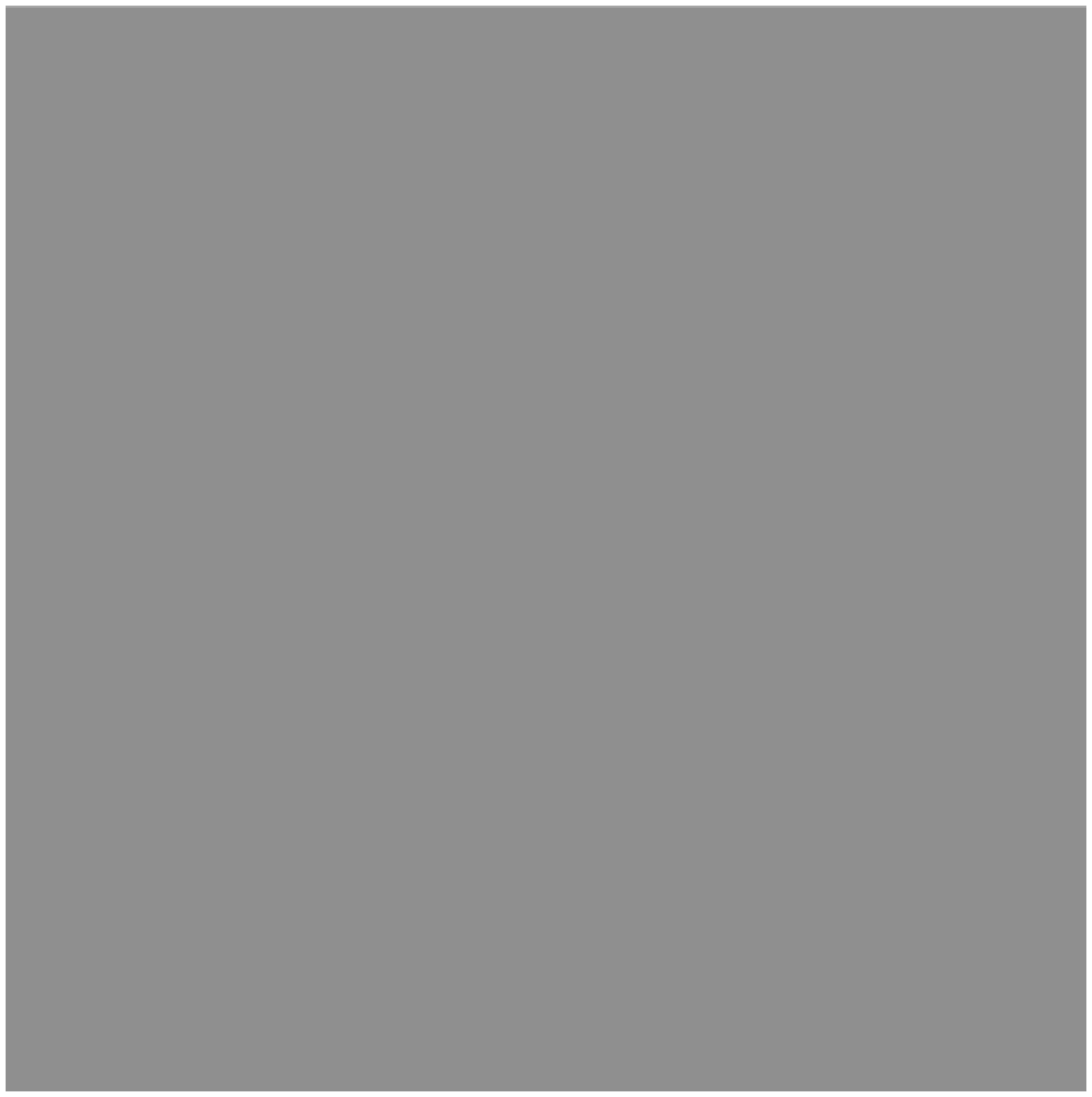} &
\includegraphics[width=3.6cm,angle=0]{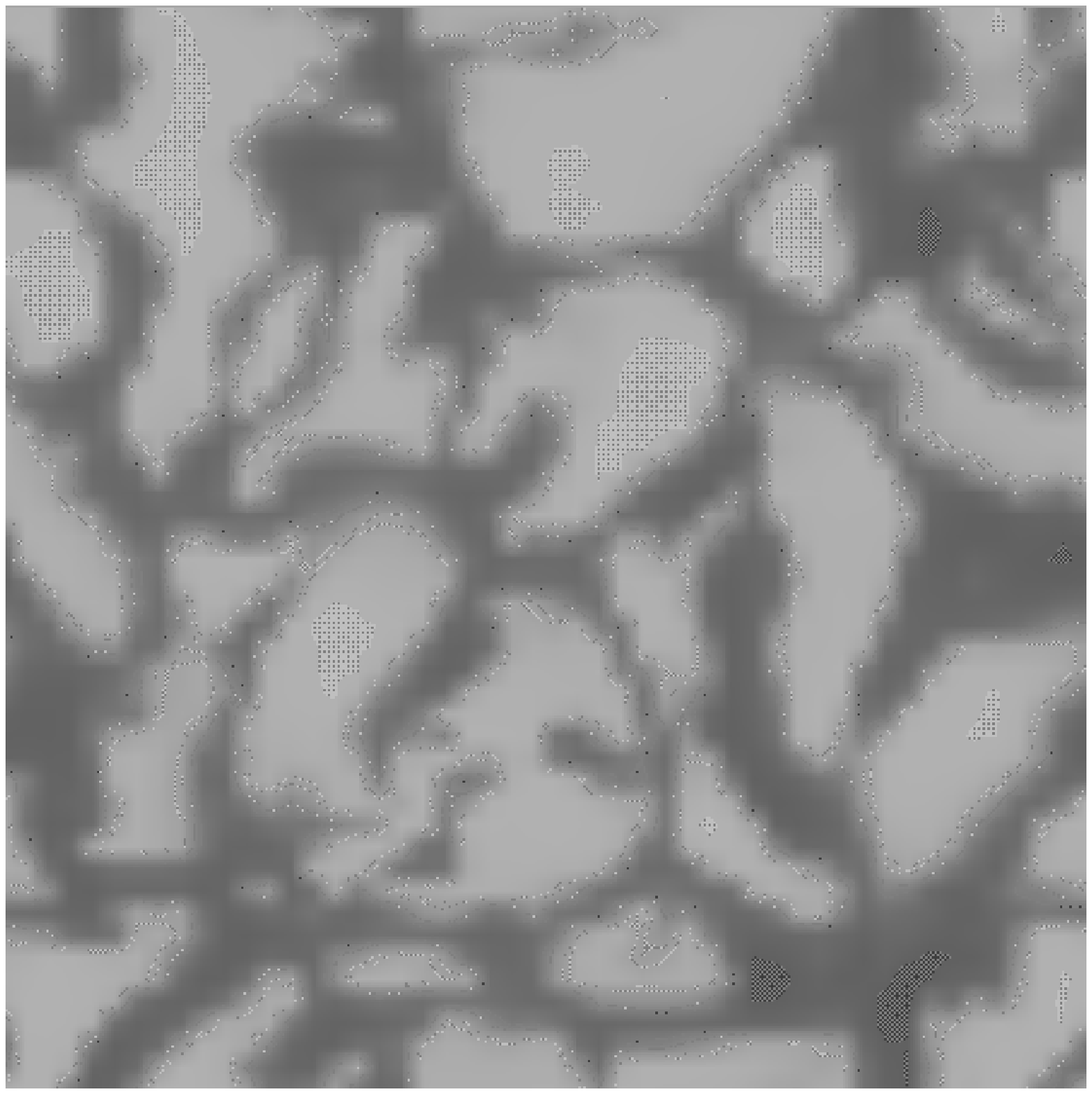} 
\end{tabular}
\protect\caption[ ]{
The gray scale indicates temperature at a given photospheric depth for a 
classical model atmosphere (left-hand panel), and in a snapshot of the
hydrodynamical simulation of Asplund et al. (2000a).
}
\label{f3}
\end{figure}

Hydrodynamical models that account for the temperature and 
velocity fields associated with convection, oscillations, and shocks
in the solar atmosphere were pioneered more than two decades ago 
(e.g., Massaguer \& Zahn 1980, Nordlund 1980, 
Hurlburt, Toomre \& Massaguer 1984) and have now reached maturity
(Stein \& Nordlund 1998, Asplund et al. 2000a).
Unlike  classical model atmospheres, which have a constant
temperature at any given depth, these models exhibit temperature and
velocity inhomogeneities that resemble solar observations.
Fig. \ref{f3} compares the temperature at a given depth in the photosphere for
 one-dimensional and three-dimensional models -- compare with Fig. \ref{f2}.

Classical static models predict perfectly symmetric spectral lines.
The line profiles produced in hydrodynamical models are asymmetric and
blue-shifted relative to their rest wavelengths as a result of the correlation
between temperature and velocity fields -- very much like the observed
line profiles (Asplund et al. 2000a, Allende Prieto et al. 2002).
In some cases, the  strengths of spectral lines predicted by  3D and 1D models 
are significantly different.
Occasionally, the ability of hydrodynamical models to reproduce closely
the observed line shapes, allows us to identify blends that otherwise would 
go unnoticed.

A recent revision of the
solar photospheric abundances of several light elements based on 3D models 
has arrived at the conclusion that both oxygen and carbon 
(but also Ne, Ar, Fe and Si) are significant less abundant, in some cases by
as much as 40 \%, than previously thought (Allende Prieto et al. 2001, 2002; 
Asplund et al. 2000b, 2004, 2005).
The revised solar abundances show that hydrodynamics (and  
departures from LTE) are not mere refinements, but crucial ingredients. 
The updated abundances are in better
agreement with other stars in the solar neighborhood, but
they also ruin an earlier excellent agreement between helioseismic 
measurements and models of the solar interior (Bahcall et al. 2005a, 
Bahcall, Serenelli \& Basu 2005c, Basu \& Antia 2004, Antia \& Basu 2005,
Montalb\'an et al. 2004) 
-- a conundrum that still needs a solution at the time of this writing
(see Bahcall, Basu \& Serenelli 2005b and Drake \& Testa 2005 for possible light 
at the end of the tunnel, or Young \& Arnett 2005 
for a different source of light).
Unfortunately, hydrodynamical models for stars other than the Sun
are not widely available (and mostly non-existent). Moreover, 
the tools available to calculate spectra from time-dependent 3D hydrodynamical
simulations are very limited, and unable, for example, to calculate
emergent absolute fluxes with realistic (line and continuum) opacities
over large spectral windows.

\begin{figure}[h!]
\centering
\includegraphics[width=7cm,angle=0]{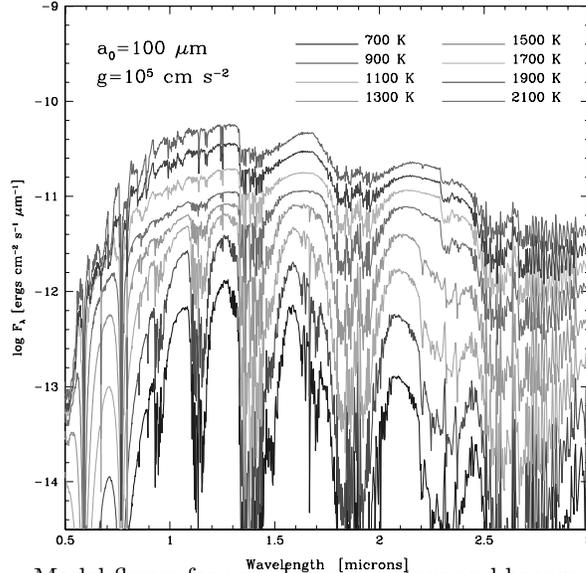} 
\protect\caption[ ]{
Model fluxes for
very low mass stars and brown dwarfs (dividing line at 
$T_{\rm eff} \sim 1700$ K)  as they
would be observed at  a distance of 10 pc 
($\log {\rm F} \propto 2 \log R + 4 \log T_{\rm eff}$, 
and the radii are adopted from
Burrows et al. 1997). Adapted from
Burrows, Sudarsky \& Hubeny 2005.
}
\label{f4}
\end{figure}

For stars with $T_{\rm eff} < 4000$ K, molecules 
become very important both in the equation of state and in  
the radiative opacity (mainly H$_2$, H$_2$O, CH$_4$, CO, N$_2$, NH$_3$, 
FeH, CrH, TiO, and VO). In addition, for atmospheres cooler than 
$T_{\rm eff} \sim 2500 $K,  solid particles, such as silicates, 
which form clouds, are also an important source of opacity, 
with alkalis as the only atoms that still contribute significantly. 
The complicated opacities pose a challenge for 
modeling the coolest stars and brown dwarfs. Handling
the formation and rainout of condensate clouds further complicates  
matters (e.g. Burrows, Sudarsky \& Hubeny 2005, Tsuji 2005).  
Fig. \ref{f4} illustrates the change in the emergent flux 
between 700 K and 2100 K for models
recently computed by Burrows et al. (2005) for a particular 
surface gravity, particle size, and cloud shape.

\subsection{Spectrum formation}
\label{spectrum}

Although the radiation field needs to be continuously evaluated 
as model construction proceeds, the detailed calculation of the emergent
spectrum for comparison with observations is usually performed afterwards. 
At this time, higher spectral resolution  and a more accurate  
modeling of the line profiles becomes affordable as the equation of
radiative transfer needs to be solved just a few times. 
It is also possible 
to estimate departures from LTE for trace species, assuming that the
atmospheric structure (e.g. temperature and electron density) 
is given by an LTE model atmosphere previously calculated.

\begin{figure}[ht!]
\centering
\includegraphics[width=7cm,angle=90]{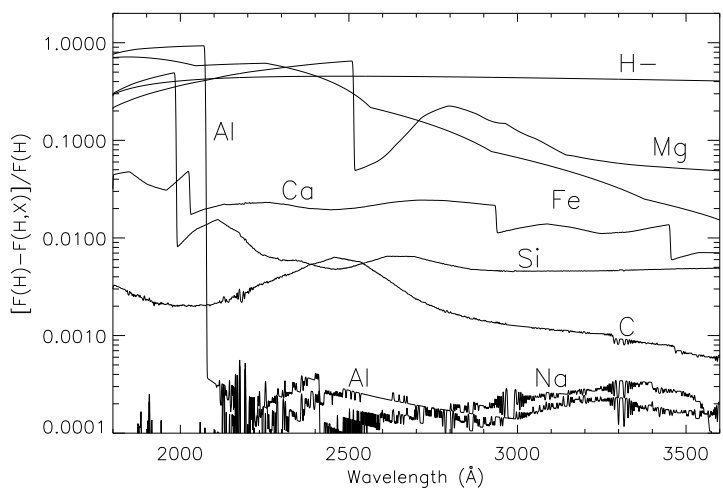}  
\protect\caption[ ]{
Relative changes in the emergent flux from a solar-like atmosphere when
continuum opacity from H$^{-}$ and several metals are added to the
atomic hydrogen opacity (Allende Prieto, Hubeny \& Lambert 2003b). 
}
\label{f5}
\end{figure}

Reliable opacities, as we have discussed in \S \ref{models}, can be
the limiting factor for obtaining realistic model atmospheres. For stars
like the Sun, the optical and infrared continuum is shaped by bound-free
and free-free absorption of H$^{-}$, a trace species nonetheless. 
In the UV window, however, 
atomic metal opacity (mainly due to Mg, Fe, and Al) becomes relevant, 
and even dominant at wavelengths below 250 nm. Quantum mechanical 
calculations in the context of
the Opacity Project (OP) 
have provided accurate photoionization cross-sections for light elements 
(Seaton 2005), and an extension of the calculations to iron ions is 
ongoing within the Iron Project (IP; Nahar \& Pradhan 2005). 
These data are (slowly) being incorporated in the calculation of
theoretical spectra (Cowley \& Bautista 2003, Allende Prieto et al. 2003a,b).
Fig. \ref{f5} illustrates the impact of the
bound-free metal absorption on the UV solar flux.

\begin{figure}[t!]
\centering
\includegraphics[width=11cm,angle=0]{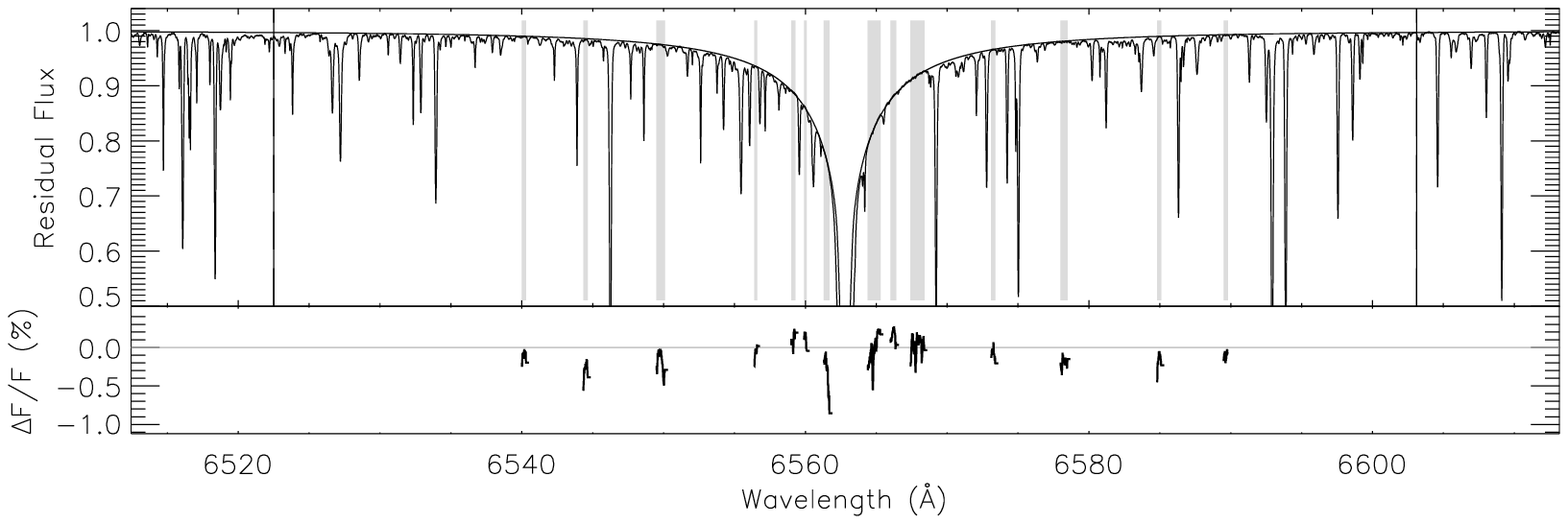}  
\includegraphics[width=11cm,angle=0]{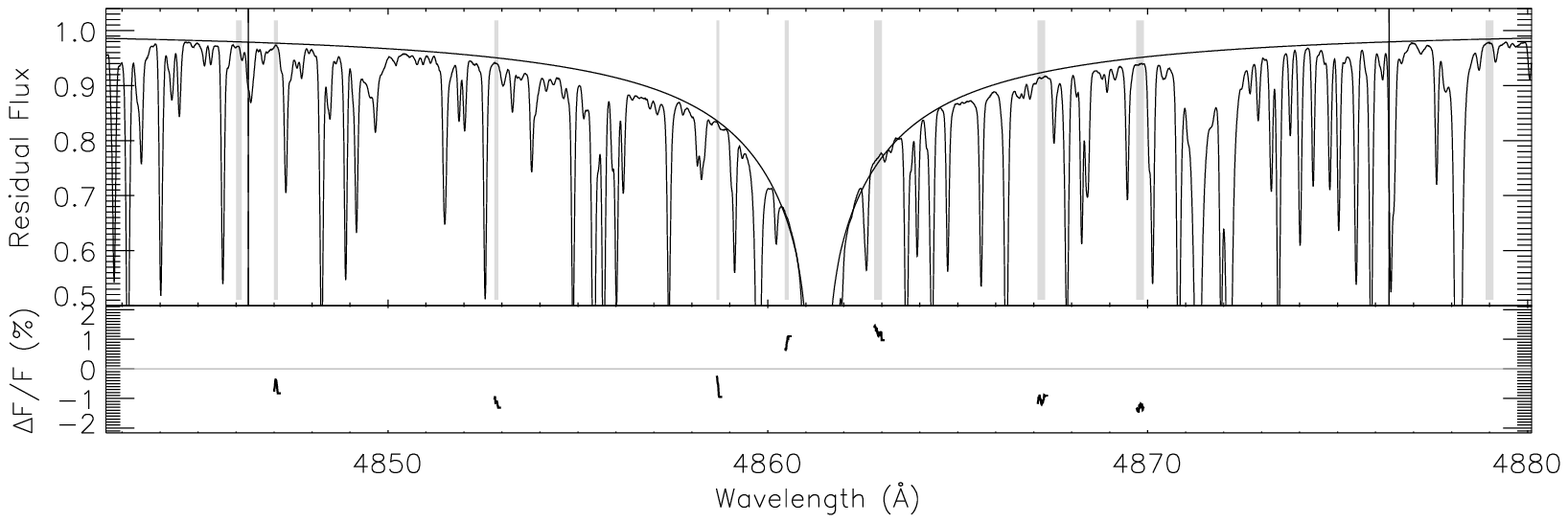}  
\protect\caption[ ]{
Fittings to the solar H$\alpha$ and H$\beta$ line profiles with a
MARCS model (Gustafsson et al. 1975, 2003) for 
a mixing-length parameter $\alpha = l/H_p = 0.5$, the Stark
broadening profiles of Stehl\'e (1994) and the hydrogen collision broadening
profiles of Barklem, Piskunov,\& O'Mara (2000). The grey areas indicate the
spectral windows used to evaluate the goodness-of-fit. 
Adapted from Barklem et al. (2002).
}
\label{f6}
\end{figure}

Another important aspect of the calculation of stellar spectra
that has been recently improved is related to the absorption line coefficients.
Improved  calculations of cross-sections for line broadening  
by elastic hydrogen collisions have been presented by Paul Barklem and 
collaborators (see Barklem, Anstee, \&  O'Mara 1998, 
Barklem \& Aspelund-Johansson 2005, and references therein). Improved absorption
coefficients are also now available for Balmer lines
(see, e.g., Barklem, Piskunov \& O'Mara 2000, Barklem et al. 2002). 
The wings of the Balmer lines, formed in deep layers very close to LTE
conditions, 
are very sensitive indicators of the effective temperature of a star
(see Fig. \ref{f6}).
Interestingly enough, the mixing-length parameter  for convection ($\alpha$) 
needs to be reduced  to about half a pressure scale height in order to get 
consistent temperatures from H$\alpha$ and H$\beta$ for solar like stars
-- the usual values inferred from standard solar models and 
multidimensional simulations of solar surface convection lead to 
$2 < \alpha < 2.1$ (Basu, Pinsonneault, \& Bahcall 2000, Robinson et al. 2004).
The new calculations of collisional broadening by atomic H have the largest 
impact for metal-poor stars, where gas pressure is higher than at solar 
metallicity, and have made it possible to reconcile effective temperatures 
inferred from colors and Balmer lines. 

The OP data 
base\footnote{http://vizier.u-strasbg.fr/topbase/home.html}, TOPBASE, includes
other useful data in addition to photoionization cross-sections: energy
levels and  transition probabilities, as well as tools to 
compute opacities and radiative forces. The computed energies for atomic levels
are not very accurate, and although the opacities include corrections for the
ground level of each species, they can be improved by using observed energies, 
readily available from other sources such as NIST\footnote{www.nist.gov}. OP and IP 
calculations assume LS coupling and neglect fine structure. For the line
opacity, one commonly uses accurate observed line wavelengths.
The most comprehensive  atomic and molecular line lists have been compiled
by Kurucz and collaborators 
(e.g., Kurucz 1993\footnote{http://kurucz.harvard.edu}).
Kurucz has also calculated and compiled
wavelengths and weights for hyperfine structure ({\it hfs}) and isotopic components.

\begin{figure}[ht!]
\centering
\includegraphics[width=9cm,angle=0]{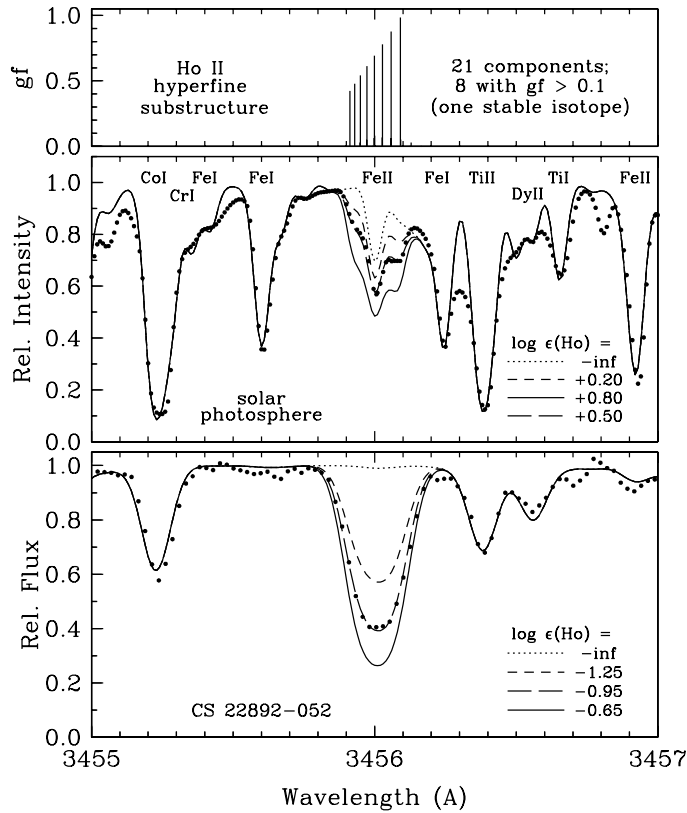}  
\protect\caption[ ]{
Ho II line at $\lambda3456$.
The lines in the top panel show the strengths
and positions of the hyperfine splitting (hfs)
components due to $^{165}$Ho, the only stable
isotope of holmium. The middle and lower panels show the spectrum of the Sun
and the metal-poor star 
CS 22892-052 (see Section \ref{golden}), respectively. The impact of
hfs in the line shape is quite dramatic, and Ho abundances inferred would be 
seriously in error if neglected. Adapted from Lawler, Sneden \& Cowan (2005).
}
\label{f7}
\end{figure}

A noteworthy development is the recent availability of 
laboratory determinations of atomic parameters
(transition probabilities, hyperfine and isotopic splitting constants, and 
partition functions) for many rare-earth elements (e.g., Den Hartog et al. 2005;
Lawler, Sneden \& Cowan 2004; Den Hartog, Wickliffe \& Lawler 2002;
Lawler, Wyart \& Blaise 2001). Fig. \ref{f7} shows how important it is
to consider {\it hfs}  when deriving abundances 
for elements like holmium.

While LTE constitutes a very useful approximation that makes the calculation of
model atmospheres and spectra simple, it is often wrong. 
Calculations of Non-LTE line formation in LTE model atmospheres for trace
species show abundant examples where line strengths interpreted 
with equilibrium (LTE) populations lead to large systematic errors 
in the inferred abundances. 
NLTE models for hot stars are now available, 
but the same cannot be said for late-type stars.
This shortcoming is most likely the result of the extended belief 
that departures from LTE are small in late-type stars.
In fact, recent calculations by Short \& Hauschildt (2005) suggest that 
there are significant changes in the temperature structure
of a solar model when LTE is relaxed. 
The same investigation also finds that a NLTE 
solar model matches the observed fluxes worse than an LTE model 
-- a puzzling result which suggests that this is still an unfinished business.

A more extensive discussion of some of the issues mentioned in this 
and the previous
section can be found in the reviews by  Gustafsson
\& J{\o}rgensen (1994), Allard et al. (1997), Asplund (2005), and
Kirkpatrick (2005).

\subsection{Observation and analysis}

As modeling improves, so do some aspects of the data acquisition and analysis
process. Quantities that have seen constant progress include the telescopes'
diameter, the spectral coverage of spectrographs, and their multiplexing
capabilities.
However, a victim of the difficulty to couple large diameter
telescopes with narrow slits, the resolving power has
been neglected. Investigations of spectral line shapes, 
which are poorly matched
by hydrostatic models and therefore nobody wants to see, 
have almost become faux pas.

Data reduction is done in most cases with packages such as IRAF or MIDAS. 
Pipelines for automatic reduction have been implemented for some
facilities and large projects (e.g. Ritter \& Washuettl 2004). 
For the most part, spectra are usually reduced 
interactively, which causes two problems: i) individual scientists, 
trying to move on as quickly as possible to the data
analysis, do not dedicate the necessary amount of time to make sure the 
reduction is the best possible, and ii) the knowledge on data reduction 
is not encapsulated into software. A more efficient and better scheme 
would involve dedicated research on the data reduction process specific 
for an instrument, with the results subsequently captured into a pipeline. 
For example, Piskunov \& Valenti  (2002) have recently 
reminded us that going the extra mile pays off.

As data sets grow in size -- more stars and more frequencies per star--
it becomes necessary to automate not only the data reduction, but also
the data analysis.
For example, a
spectroscopic analysis typically starts with a search for the best estimates 
of the atmospheric parameters: effective temperature, surface gravity and
overall metal content. Such step is amenable to automation 
(Katz et al. 1998, Snider et al. 2001, Allende Prieto 2003, 
Erspamer \& North 2003, Willemsen et al. 2005). 
Neural networks, genetic algorithms, self-organizing
maps and many other kinds of optimization methods can help to relief
the burden on the busy astronomer.

\section{Stellar abundances. What for?}
\label{whatfor}

After some hard work, we get to the most exciting part: what can we do with 
the derived chemical abundances? A possible criterion to classify the 
applications of stellar abundances  is whether or not they are based on
the {\it golden rule}. The golden rule, a term introduced in this context 
by David Lambert, states that

{\it The surface composition of a star reflects that of the interstellar
medium at the time and location where the star formed.}

As with any other good-looking law, the golden rule is often broken.
But we will first see how to take advantage of those cases when it 
seems to hold.

\subsection{When the {\bf golden rule} applies}
\label{golden}

Only  $^1$H, $^2$H, $^3$He, $^4$He,
$^7$Li and $^6$Li were produced in significant amounts in the big bang, 
while heavier nuclei are mostly 
the result of stellar nucleosynthesis. When  stars explode as supernovae,
or lose mass from their envelopes  by the action of stellar winds or
thermal pulses on the asymptotic  giant branch (AGB), 
the metals produced by stellar nucleosynthesis 
(explosive or not) enrich the interstellar medium.
As low-mass stars have very long life spans (the hydrogen burning
time scales $\propto M^{-5/2}$), if the golden rule applies, one can
use stars of different ages to track the metal enrichment of the
interstellar medium with time. An account of the state of this field
is given in this volume by Yeshe Fenner, so I just briefly
mention some recent observational developments that illustrate how
stellar abundances can help to understand how galaxies form.

\begin{figure}[ht!]
\centering
\includegraphics[width=8cm,angle=0]{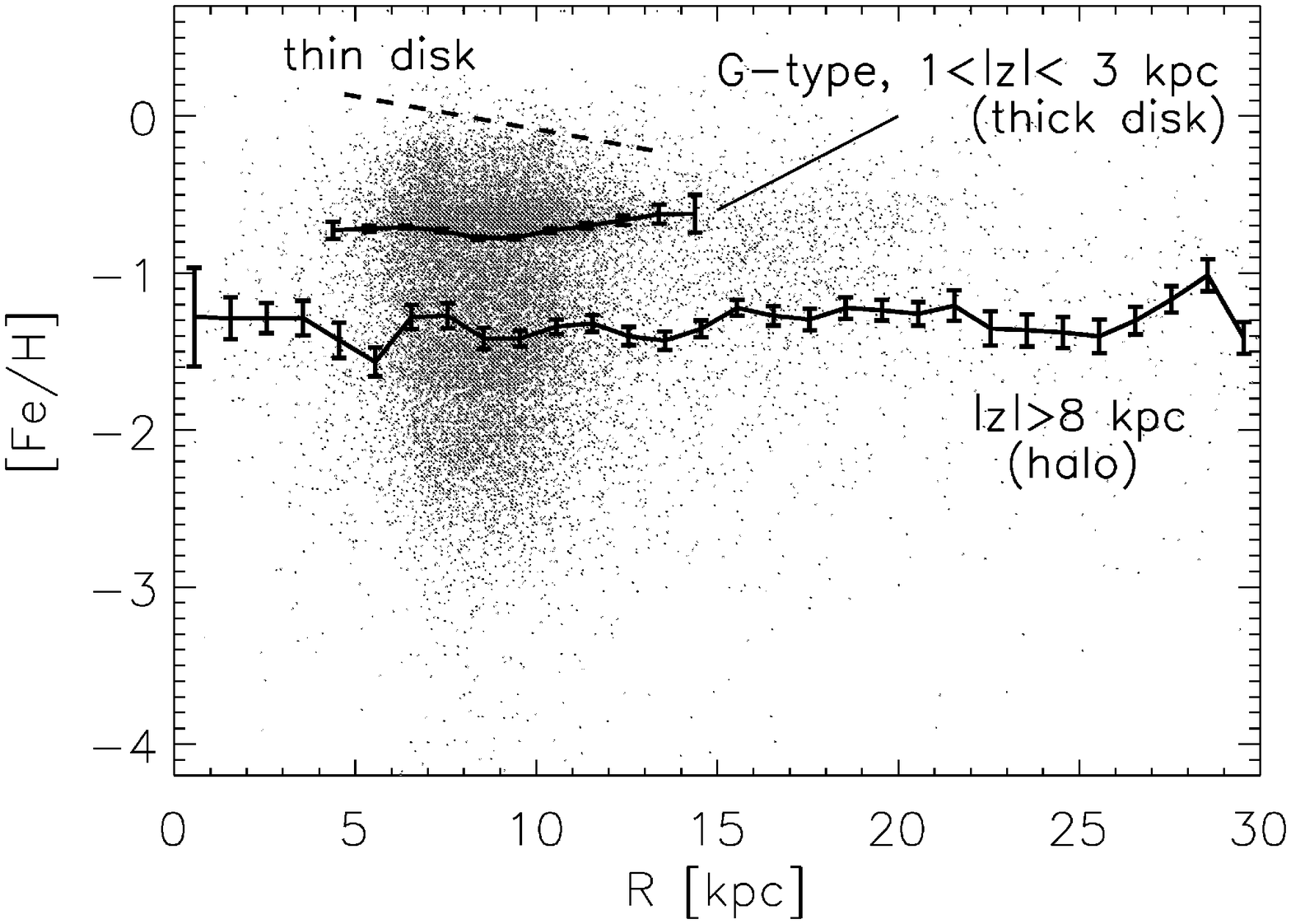}  
\includegraphics[width=9cm,angle=0]{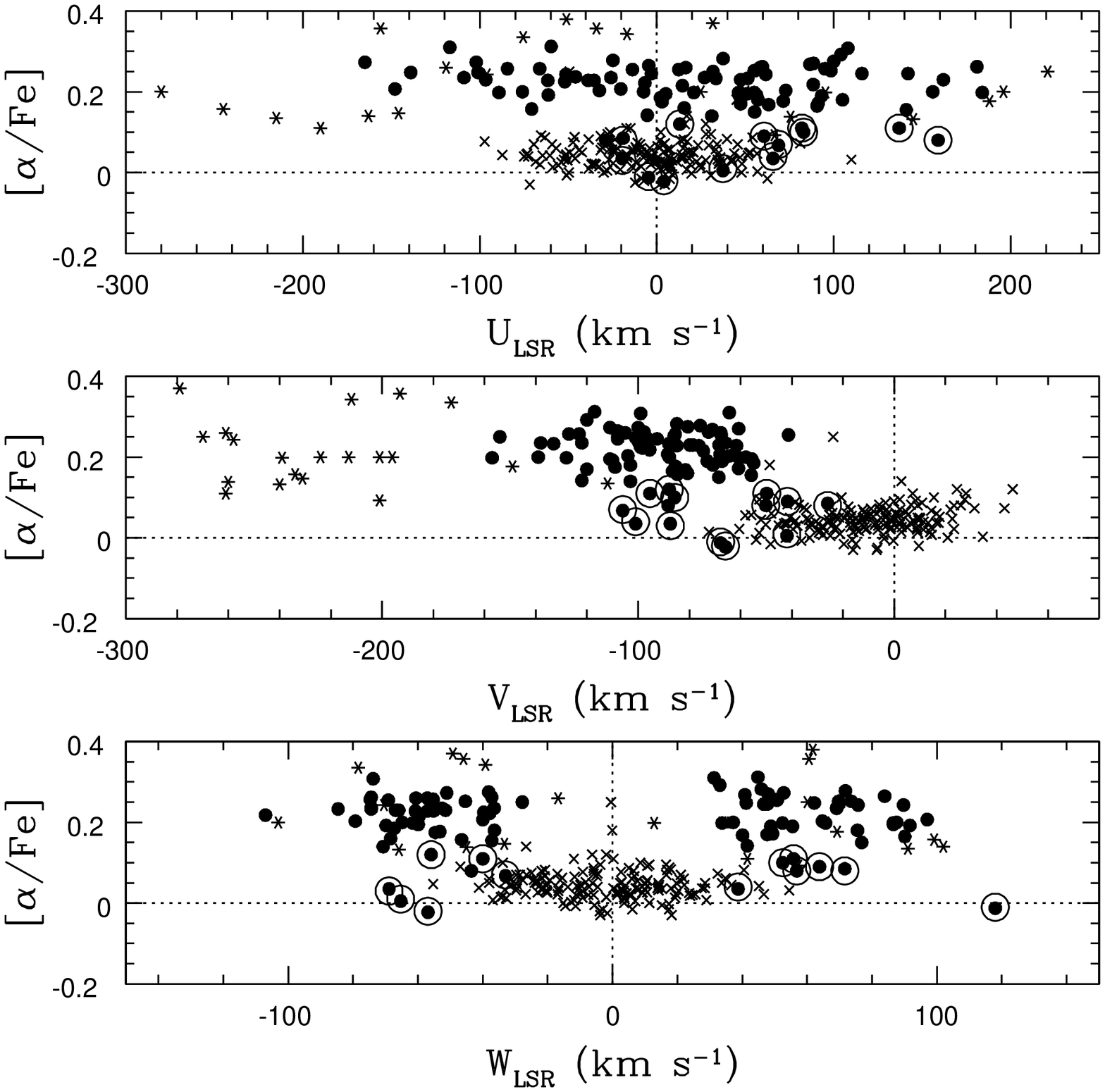}   
\protect\caption[ ]{
Top panel: some 20,000 F- and G-type stars with $14<V< 20$ mag 
spectroscopically observed as part of the SDSS are used here to 
trace the median metallicity of the thick disk and the halo as a function
of distance from the Galactic center ($R$ in cylindrical coordinates). 
In this figure z is the distance from the galactic plane.
The flat values of the median 
iron abundances contrast with the gradients reported in the literature
for the thin disk (the dash line shows the mean metal abundances derived 
for B-type stars by Daflon \& Cunha 2004). Adapted from Allende Prieto 
et al. (2006). Bottom panels:
Chemical and kinematical separation of thick disk (filled circles),
 thin disk (crosses), and halo (asterisks) stars -- plus
the enigmatic stars with thick-disk kinematics and thin-disk abundances 
(filled circles surrounded by open circles). Here $\alpha$ indicates  an
average of Mg, Si, Ca and Ti. Adapted from Reddy et al. (2006).
}
\label{f8}
\end{figure}

The disk of the Milky Way was first found to possess a {\it second} 
component with a larger scale height 
by Gilmore \& Reid (1983). This
component, which is known as the {\it thick disk}, contains more metal-poor and
older stars than the thin disk. Thick disk stars are not only chemical, 
but also kinematically distinct from the thin disk population: a larger
velocity ellipsoid, and an average galactic rotation velocity 
that lags the thin disk by some 30--40 km s$^{-1}$, as illustrates Fig. \ref{f8}
(see also Gratton et al. 1996; Fuhrmann 1998; Bensby et al. 2005; 
Mishenina et al. 2004; Reddy et al. 2003).
A recent study that exploits calibration stars observed as part of the
Sloan Digital Sky Survey (SDSS) suggests that the peak of 
the metallicity distribution of the thick disk is always at [Fe/H] $\sim -0.7$ 
between 4 and 14 kpc from the galactic center 
(Allende Prieto et al. 2006), 
while local thick disk members selected based on 
kinematics indicate a high chemical homogeneity in this population
(Reddy, Lambert \& Allende Prieto 2006). 
These and other observations
seem consistent with the thick disk emerging in a period of
intensive accretion of  smaller gas-rich galaxies by the Milky Way 
at a redshift $> 1$ or, equivalently, more than $8\times 10^9$ yr 
ago (Brook et al. 2005).

Some of the oldest stars in the Milky Way have extremely low metal
abundances, much lower than any globular cluster in the Galaxy. Their
compositions may resemble that of the early Galaxy,
with Li/H ratios similar to the proportions produced in the
big bang. This, in fact, was the interpretation given
by Spite \& Spite (1982) to the nearly constant values of Li/H found
in the surface of turn-off field stars with very low metal abundances.
Experts still debate whether or not the
Li abundance is exactly the same in the most metal-poor turn-off stars 
(e.g., Ryan et al. 1999, Mel\'endez \& Ram\'{\i}rez 2004), 
but for our purposes it suffices to say that big bang nucleosynthesis
models yield different Li/H ratios depending on the universe's photon
to baryon ratio, and therefore the  lithium abundances
in stars can constrain such an important cosmological parameter.

\begin{figure}[ht!]
\centering
\includegraphics[width=8cm,angle=0]{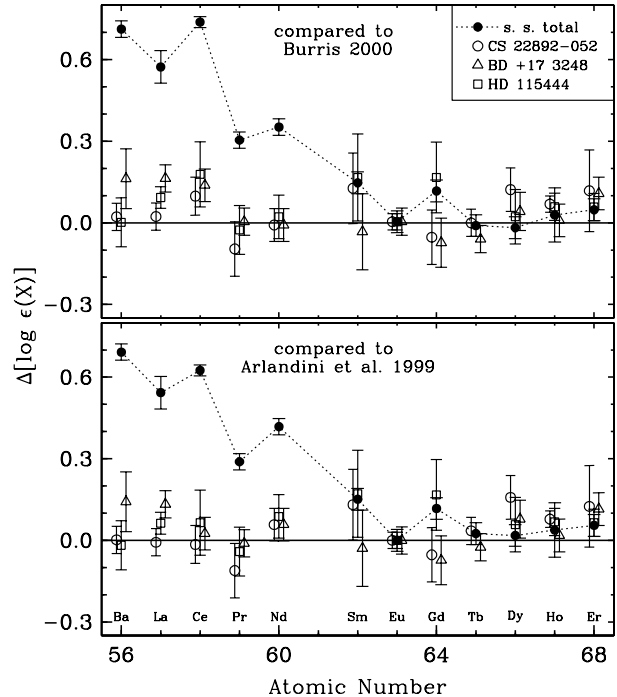}  
\protect\caption[ ]{
Differences between the abundances of heavy neutron capture elements
in three stars and the {\it r-process} component of the 
solar abundances. The {\it r-process} solar abundances were inferred by
subtracting the {\it s-process} (slow neutron capture) abundances estimated
by two different methods: an empirical approach (Burris et al. 2000) or
AGB stars nucleosynthesis models (Arlandini et al. 1999). The patterns have
been normalized to force a perfect agreement for europium. Adapted from
Lawler et al. (2005).
}
\label{f9}
\end{figure}

When old very metal-poor halo stars formed, there were very few metals
around to be incorporated in the stars' atmospheres. As massive stars
die exploding as Type II supernovae, there must have been a period
of high chemical heterogeneity in the halo. If a star happened to form
at such early stages near the location where a supernova exploded, it could
pick up the proportions of heavy elements produced before and during 
the explosion. That may have been the case for  CS 22892-052, an
extremely metal-poor ([Fe/H]$=-3.1$) and old (age $>11\times 10^9$ yr) 
halo giant  which is highly enriched in neutron-capture elements 
(Sneden et al. 1994, 2003), and also for several other well-known cases, such
as BD $+17 ~3248$ (Cowan et al. 2002), or HD 115444 (Westin et al. 2000). 
As Fig. \ref{f9} illustrates, 
the abundance patterns of heavy neutron-capture elements in these stars
are remarkably similar to the {\it r-process} contribution to the 
solar-system abundances of these elements (as inferred by two different methods:
Burris et al. 2000 vs. Arlandini et al. 1999). 
The {\it r-process} nucleosynthesis
operates when  high neutron fluxes are available, and it 
is usually associated with Type II supernovae. Therefore measuring 
abundances in these stars allows us to study supernova yields. Ongoing active
searches for more of these interesting objects are already producing
results (Barklem et al. 2005).

\subsection{If the {\bf golden rule} does not apply}

The chemical composition of the atmosphere of a star may have changed 
since the star formed. 
As research has already shown, this can happen by a myriad of different 
processes.
An interesting example of the use of stellar abundances 
under these circumstances is very close to home. 
Chemical abundances derived for the solar photosphere have
been found to provide valuable clues on the structure of the solar interior.
Considering diffusion in the solar convection zone 
is necessary to match the p-mode frequencies from  helioseismic observations 
(e.g., Demarque \& Guenther 1988, Christensen-Dalsgaard 2002, Lodders 2003). 
Since the formation of the Sun 4.57$\times 10^9$ years ago, 
the surface abundance of He relative to H has  seen a reduction of 
18 \%, while the abundances of heavier elements have been reduced by 
16 \%. 

The surface abundances of Li or Be in late-type stars can change 
significantly when the material is transported by convection 
to more interior zones where
the temperature reaches a few million degrees and these light nuclei 
are destroyed. 
This violation of the golden rule has been exploited 
to learn about stellar structure by studying 
abundances in clusters (e.g. Boesgaard et al. 2004, 
Garc\'{\i}a L\'opez, Rebolo \& P\'erez de Taoro 1995).

We discussed above how supernova yields can be inferred from rare
metal-poor stars when the golden rule applies, but some have also attempted 
to do the same when the rule is not respected. Binary systems that have as a 
primary a neutron star or a black hole offer a chance to detect supernova 
ejecta that may have been blocked by the companion. 
Gonz\'alez-Hern\'andez et al. (2005) detected unusual enhancements of
Ti and Ni ratios in the K-type companion of the neutron star Cen X-4,
and concluded by comparison with theoretical calculations 
that the observed abundances could be explained by a
spherically symmetric supernova explosion.
Nevertheless, the kinematics of the system supports an asymmetric supernova, 
leaving open an interesting puzzle.

\begin{figure}[t!]
\centering
\includegraphics[width=6cm,angle=0]{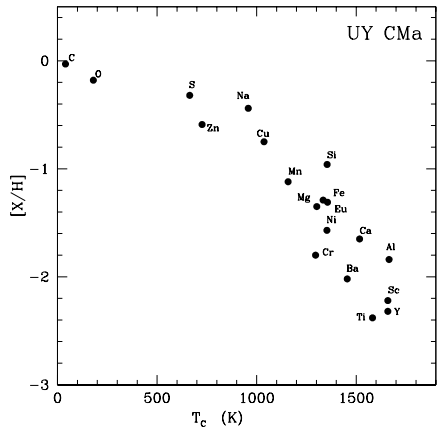}  
\protect\caption[ ]{
Abundance pattern in the RV Tauri star UY CMa. There is a strong
correlation of the abundances with the condensation temperature.
Adapted from Giridhar et al. (2005).
}
\label{f10}
\end{figure}

Perhaps one of the worse cases of the golden rule being broken is that of the
RV Tauri variables. The photospheres of these stars exhibit dramatic
abundance anomalies that correlate with condensation temperature (T$_{\rm c}$). 
Elements with a high T$_{\rm c}$ 
can be depleted by several orders of magnitude respect to those with
a low T$_{\rm c}$. The elements depleted form 
dust grains, ending up in some sort of circumstellar reservoir, o being 
driven away by a stellar wind. Fig. \ref{f10} shows the abundance pattern 
derived for UY Cam by Giridhar et al. (2005). 
Metal-deficient post-AGB stars such as 
HR 4049 (Van Winckel, Waelkens \& Waters 1995) exhibit
a photospheric iron abundance [Fe/H] $\simeq -5.0$ but most likely 
started up with a near-solar Fe/H ratio. HR 4049 and its class seem
to be all members of binary systems, and the same  could be true for 
RV Tauris.  We refer the reader to the recent review by Van Winckel (2003)
for more on the amazing fauna of post-AGB stars.

\section{Conclusions}

Recent noteworthy trends in research on stellar abundances 
include the development
of improved model atmospheres for solar-like stars considering hydrodynamics,
the availability of full-NLTE line-blanketed model atmospheres for hot stars,
significant improvements in the opacities and the treatment of clouds for
the coolest stars and brown dwarfs,
the emergence of large spectroscopic surveys producing public data
bases (e.g., SDSS, Elodie, S$^4$N, SEGUE, GALEX), 
significant progress toward the automation of spectroscopic analyses,
and in availability of accurate atomic data for neutron-capture elements.

A desirable short-term future would include
3D model atmospheres for late-type stars of all types, 
1D models in full NLTE, NLTE line formation in 3D models, 
solar atlases observed at different positions on the disk to test
line formation and model atmospheres, 
stronger efforts to measure/compute the necessary atomic and molecular data, 
stronger efforts to use the newly available atomic and molecular data, 
full analysis automation, 
a new generation of all-in-one archives (or virtual observatories)
 for both multi-wavelength observations and models, 
and ultra-high resolution spectrographs on large telescopes.
Let's make it happen!

\acknowledgements I am thankful to the organizers for a delightful meeting,
and the opportunity to think (loud) about the field of stellar abundance 
determination. Some of the figures in this paper 
were kindly provided by A. Burrows, S. Giridhar, 
I. Hubeny, T. Lanz, J. Lawler, B. Reddy, and C. Sneden.
The Swedish 1-m Solar Telescope is operated on the island of 
La Palma by the Institute for Solar Physics of the Royal Swedish 
Academy of Sciences in the Spanish Observatorio del 
Roque de los Muchachos of the Instituto de Astrof\'{\i}sica de Canarias.
Support from NASA (NAG5-13057, NAG5-13147) is gratefully acknowledged.


\begin{thebibliography}{}


\bibitem[Allard et al.(1997)]{1997ARA&A..35..137A} Allard, F., Hauschildt, 
P.~H., Alexander, D.~R., \& Starrfield, S.\ 1997, \araa, 35, 137 
 
\bibitem[Allende Prieto(2003)]{2003MNRAS.339.1111A} Allende Prieto, C.\ 
2003, \mnras, 339, 1111  
 
\bibitem[Allende Prieto et al.(2002)]{2002ApJ...567..544A} Allende Prieto, 
C., Asplund, M., Garc\'{\i}a L{\'o}pez, R.~J., \& Lambert, D.~L.\ 2002, \apj, 567, 
544 

\bibitem[]{} Allende Prieto, C., Beers, T. C., Wilhelm, R., Newberg, H. J.,
Rockosi, C. M., Yanny, B., \& Lee, Y. S. 2006, ApJ, 636, 804

\bibitem[Allende Prieto et al.(2001)]{2001ApJ...556L..63A} Allende Prieto, 
C., Lambert, D.~L., \& Asplund, M.\ 2001, \apjl, 556, L63 

\bibitem[Allende Prieto et al.(2002)]{2002ApJ...573L.137A} Allende Prieto, 
C., Lambert, D.~L., \& Asplund, M.\ 2002, \apjl, 573, L137 
 

\bibitem[Allende Prieto et al.(2003)]{2003ApJS..147..363A} Allende Prieto, 
C., Lambert, D.~L., Hubeny, I., \& Lanz, T.\ 2003a, \apjs, 147, 363 

\bibitem[Allende Prieto et al.(2003)]{2003ApJ...591.1192A} Allende Prieto, 
C., Hubeny, I., \& Lambert, D.~L.\ 2003b, \apj, 591, 1192 


\bibitem[Antia \& Basu(2005)]{2005ApJ...620L.129A} Antia, H.~M., \& Basu, 
S.\ 2005, \apjl, 620, L129 

\bibitem[Arlandini et al.(1999)]{1999ApJ...525..886A} Arlandini, C., 
K{\"a}ppeler, F., Wisshak, K., Gallino, R., Lugaro, M., Busso, M., \& 
Straniero, O.\ 1999, \apj, 525, 886 
 

\bibitem[Asplund(2005)]{2005ARA&A..43..481A} Asplund, M.\ 2005, \araa, 43, 
481 
 
\bibitem[Asplund et al.(2004)]{2004A&A...417..751A} Asplund, M., Grevesse, 
N., Sauval, A.~J., Allende Prieto, C., \& Kiselman, D.\ 2004, \aap, 417, 
751 

\bibitem[Asplund et al.(2005)]{2005A&A...431..693A} Asplund, M., Grevesse, 
N., Sauval, A.~J., Allende Prieto, C., \& Blomme, R.\ 2005, \aap, 431, 693   

\bibitem[Asplund et al.(2000)]{2000A&A...359..729A} Asplund, M., Nordlund, 
{\AA}., Trampedach, R., Allende Prieto, C., \& Stein, R.~F.\ 2000a, \aap, 
359, 729 

\bibitem[Asplund et al.(2000)]{2000A&A...359..743A} Asplund, M., Nordlund, 
{\AA}., Trampedach, R., \& Stein, R.~F.\ 2000b, \aap, 359, 743 
 

\bibitem[Bahcall et al.(2005)]{2005ApJ...618.1049B} Bahcall, J.~N., Basu, 
S., Pinsonneault, M., \& Serenelli, A.~M.\ 2005a, \apj, 618, 1049 

\bibitem[Bahcall et al.(2005)]{2005ApJ...631.1281B} Bahcall, J.~N., Basu, 
S., \& Serenelli, A.~M.\ 2005b, \apj, 631, 1281 
 
\bibitem[Bahcall et al.(2005)]{2005ApJ...621L..85B} Bahcall, J.~N., 
Serenelli, A.~M., \& Basu, S.\ 2005c, \apjl, 621, L85 

\bibitem[Barklem et al.(1998)]{1998PASA...15..336B} Barklem, P.~S., Anstee, 
S.~D., \& O'Mara, B.~J.\ 1998, PASA, 15, 336 

\bibitem[Barklem \& Aspelund-Johansson(2005)]{2005A&A...435..373B} Barklem, 
P.~S., \& Aspelund-Johansson, J.\ 2005, \aap, 435, 373 

\bibitem[Barklem et al.(2005)]{2005A&A...439..129B} Barklem, P.~S., et al.\ 
2005, \aap, 439, 129 

\bibitem[Barklem et al.(2000)]{2000A&A...363.1091B} Barklem, P.~S., 
Piskunov, N., \& O'Mara, B.~J.\ 2000, \aap, 363, 1091 
 
\bibitem[Barklem et al.(2002)]{2002A&A...385..951B} Barklem, P.~S., 
Stempels, H.~C., Allende Prieto, C., Kochukhov, O.~P., Piskunov, N., \& 
O'Mara, B.~J.\ 2002, \aap, 385, 951 
 
  
\bibitem[Basu \& Antia(2004)]{2004ApJ...606L..85B} Basu, S., \& Antia, 
H.~M.\ 2004, \apjl, 606, L85 

\bibitem[Basu et al.(2000)]{2000ApJ...529.1084B} Basu, S., Pinsonneault, 
M.~H., \& Bahcall, J.~N.\ 2000, \apj, 529, 1084 
 
\bibitem[]{} Bensby, T., Feltzing, S., Lundstr\"{o}m, I., \&
 Ilyin, I. 2005, A\&A, 433, 185
 
\bibitem[Boesgaard et al.(2004)]{2004ApJ...613.1202B} Boesgaard, A.~M., 
Armengaud, E., King, J.~R., Deliyannis, C.~P., \& Stephens, A.\ 2004, \apj, 
613, 1202 
  
 
\bibitem[]{} Brook, C. B., Veilleux, V., Kawata, D., Martel, H., \& 
Gibson, B. K. 2005b, Proceedings of "Island Universes: 
Structure and Evolution of Disk Galaxies" (astro-ph/0511002)
 
 
\bibitem[B{\"o}hm-Vitense(1958)]{1958ZA.....46..108B} B{\"o}hm-Vitense, E.\ 
1958, Zeitschrift fur Astrophysics, 46, 108 

\bibitem[Burris et al.(2000)]{2000ApJ...544..302B} Burris, D.~L., 
Pilachowski, C.~A., Armandroff, T.~E., Sneden, C., Cowan, J.~J., \& Roe, 
H.\ 2000, \apj, 544, 302 

\bibitem[Burrows et al.(1997)]{1997ApJ...491..856B} Burrows, A., et al.\ 
1997, \apj, 491, 856 

\bibitem[Burrows et al.(2005)]{2005astro.ph..9066B} Burrows, A., Sudarsky, 
D., \& Hubeny, I.\ 2005, ApJ, submitted
(astro-ph/0509066)
 
\bibitem[Christensen-Dalsgaard(2002)]{2002RvMP...74.1073C} 
Christensen-Dalsgaard, J.\ 2002, Reviews of Modern Physics, 74, 1073 

\bibitem[Cowan et al.(2002)]{2002ApJ...572..861C} Cowan, J.~J., et al.\ 
2002, \apj, 572, 861 

\bibitem[Cowley \& Bautista(2003)]{2003MNRAS.341.1226C} Cowley, C.~R., \& 
Bautista, M.\ 2003, \mnras, 341, 1226 

\bibitem[Daflon \& Cunha(2004)]{2004ApJ...617.1115D} Daflon, S., \& Cunha,
K.\ 2004, \apj, 617, 1115

\bibitem[Demarque \& Guenther(1988)]{1988IAUS..123...91D} Demarque, P., \& 
Guenther, D.~B.\ 1988, IAU Symp.~123: Advances in Helio- and 
Astero-seismology, 123, 91 
 
 
\bibitem[Den Hartog et al.(2005)]{2005ApJ...619..639D} Den Hartog, E.~A., 
Herd, M.~T., Lawler, J.~E., Sneden, C., Cowan, J.~J., \& Beers, T.~C.\ 
2005, \apj, 619, 639 

\bibitem[Den Hartog et al.(2002)]{2002ApJS..141..255D} Den Hartog, E.~A., 
Wickliffe, M.~E., \& Lawler, J.~E.\ 2002, \apjs, 141, 255 
 
 
\bibitem[Drake \& Testa(2005)]{2005Natur.436..525D} Drake, J.~J., \& Testa, 
P.\ 2005, Nature, 436, 525 

\bibitem[Erspamer \& North(2003)]{2003A&A...398.1121E} Erspamer, D., \& 
North, P.\ 2003, \aap, 398, 1121 

\bibitem[]{} Fuhrmann, K. 1998, A\&A, 338, 161

\bibitem[Garcia Lopez et al.(1995)]{1995A&A...302..184G} Garc\'{\i}a L\'opez, 
R.~J., Rebolo, R., \& P\'erez de Taoro, M.~R.\ 1995, \aap, 302, 184 
 
\bibitem[Gilmore \& Reid(1983)]{1983MNRAS.202.1025G} Gilmore, G., \& Reid, 
N.\ 1983, \mnras, 202, 1025 

\bibitem[Giridhar et al.(2005)]{2005ApJ...627..432G} Giridhar, S., Lambert, 
D.~L., Reddy, B.~E., Gonzalez, G., \& Yong, D.\ 2005, \apj, 627, 432 
 

\bibitem[Gonz{\'a}lez Hern{\'a}ndez et al.(2005)]{2005ApJ...630..495G} 
Gonz{\'a}lez Hern{\'a}ndez, J.~I., Rebolo, R., Israelian, G., Casares, J., 
Maeda, K., Bonifacio, P., \& Molaro, P.\ 2005, \apj, 630, 495 
   
\bibitem[Gratton et al.(1996)]{1996ASPC...92..307G} Gratton, R., Carretta, 
E., Matteucci, F., \& Sneden, C.\ 1996, ASP Conf.~Ser.~ 92: 
Formation of the Galactic Halo, H. L. Morrison and 
A. Sarajedini, eds., 92, 307 
 
\bibitem[Gustafsson et al.(1975)]{1975A&A....42..407G} Gustafsson, B., 
Bell, R.~A., Eriksson, K., \& Nordlund, \AA.\ 1975, \aap, 42, 407 

\bibitem[Gustafsson et al.(2003)]{2003ASPC..288..331G} Gustafsson, B., 
Edvardsson, B., Eriksson, K., Mizuno-Wiedner, M., J{\o}rgensen, U.~G., \& 
Plez, B.\ 2003, ASP Conf.~Ser.~288: Stellar Atmosphere Modeling, 288, 331 
 
\bibitem[Gustafsson \& Jorgensen(1994)]{1994A&ARv...6...19G} Gustafsson, 
B., \& J{\o}rgensen, U.~G.\ 1994, \aapr, 6, 19 
   
\bibitem[Hillier(2003)]{2003ASPC..288..199H} Hillier, D.~J.\ 2003, ASP 
Conf.~Ser.~288: Stellar Atmosphere Modeling, 288, 199 

\bibitem[Hubeny \& Lanz(1995)]{1995ApJ...439..875H} Hubeny, I., \& Lanz, 
T.\ 1995, \apj, 439, 875 

\bibitem[Hurlburt et al.(1984)]{1984ApJ...282..557H} Hurlburt, N.~E., 
Toomre, J., \& Massaguer, J.~M.\ 1984, \apj, 282, 557 

\bibitem[Katz et al.(1998)]{1998A&A...338..151K} Katz, D., Soubiran, C., 
Cayrel, R., Adda, M., \& Cautain, R.\ 1998, \aap, 338, 151 
 
\bibitem[Kirkpatrick(2005)]{2005ARA&A..43..195K} Kirkpatrick, J.~D.\ 2005, 
\araa, 43, 195 
 

 
\bibitem[Kurucz(1993)]{1993KurCD..15.....K} Kurucz, R.\ 1993,  CD-ROM No.~15.~Cambridge, 
Mass.: SAO
 
\bibitem[Lanz \& Hubeny(2003)]{2003ApJS..146..417L} Lanz, T., \& Hubeny, 
I.\ 2003, \apjs, 146, 417 

\bibitem[Lawler et al.(2004)]{2004ApJ...604..850L} Lawler, J.~E., Sneden, 
C., \& Cowan, J.~J.\ 2004, \apj, 604, 850 

\bibitem[Lawler et al.(2001)]{2001ApJS..137..351L} Lawler, J.~E., Wyart, 
J.-F., \& Blaise, J.\ 2001, \apjs, 137, 351 
 
\bibitem[Lodders(2003)]{2003ApJ...591.1220L} Lodders, K.\ 2003, \apj, 591, 
1220 
  
\bibitem[Massaguer \& Zahn(1980)]{1980A&A....87..315M} Massaguer, J.~M., \& 
Zahn, J.-P.\ 1980, \aap, 87, 315 

\bibitem[]{} Mishenina, T.V., Soubiran, C., Kovtyukh, V.V., \& Korotin, S.A. 2004, 
A\&A, 418, 551

\bibitem[Montalb{\'a}n et al.(2004)]{2004soho...14..574M} Montalb{\'a}n, 
J., Miglio, A., Noels, A., Grevesse, N., \& di Mauro, M.~P.\ 2004, ESA 
SP-559: SOHO 14 Helio- and Asteroseismology: Towards a Golden Future, 14, 
574 
 
\bibitem[Mel{\'e}ndez \& Ram{\'{\i}}rez(2004)]{2004ApJ...615L..33M} 
Mel{\'e}ndez, J., \& Ram{\'{\i}}rez, I.\ 2004, \apjl, 615, L33 
 

\bibitem[Nahar \& Pradhan(2005)]{2005A&A...437..345N} Nahar, S.~N., \& 
Pradhan, A.~K.\ 2005, \aap, 437, 345 
 

\bibitem[Nordlund(1980)]{1980LNP...114..213N} Nordlund, \AA.\ 1980, Lecture 
Notes in Physics, Berlin Springer Verlag, 114, 213  
  
\bibitem[]{} Pauldrach, A. W. A., Hoffmann, T. L., Lennon, M. 2001, A\&A, 375, 161

\bibitem[Piskunov \& Valenti(2002)]{2002A&A...385.1095P} Piskunov, N.~E., 
\& Valenti, J.~A.\ 2002, \aap, 385, 1095 
 
\bibitem[]{} Rauch, T. 2003, A\&A, 403, 709

\bibitem[Ryan et al.(1999)]{1999ApJ...523..654R} Ryan, S.~G., Norris, 
J.~E., \& Beers, T.~C.\ 1999, \apj, 523, 654 
 

\bibitem[]{} Reddy, B.E., Tomkin, J., Lambert, D.L., \& 
Allende Prieto, C. 2003, MNRAS, 340, 304

\bibitem[]{} Reddy, B.E., Lambert, D.L., \& Allende Prieto, C. 2006, 
MNRAS, in press

\bibitem[Repolust et al.(2004)]{2004A&A...415..349R} Repolust, T., Puls, 
J., \& Herrero, A.\ 2004, \aap, 415, 349 
 

\bibitem[Ritter \& Washuettl(2004)]{2004AN....325..663R} Ritter, A., \& 
Washuettl, A.\ 2004, Astronomische Nachrichten, 325, 663 
 

\bibitem[Robinson et al.(2004)]{2004MNRAS.347.1208R} Robinson, F.~J., 
Demarque, P., Li, L.~H., Sofia, S., Kim, Y.-C., Chan, K.~L., \& Guenther, 
D.~B.\ 2004, \mnras, 347, 1208 

\bibitem[Seaton(2005)]{2005MNRAS.362L...1S} Seaton, M.~J.\ 2005, \mnras, 
362, L1 

\bibitem[Short \& Hauschildt(2005)]{2005ApJ...618..926S} Short, C.~I., \& 
Hauschildt, P.~H.\ 2005, \apj, 618, 926 

\bibitem[Sneden et al.(1994)]{1994ApJ...431L..27S} Sneden, C., Preston, 
G.~W., McWilliam, A., \& Searle, L.\ 1994, \apjl, 431, L27 
 
\bibitem[Sneden et al.(2003)]{2003ApJ...591..936S} Sneden, C., et al.\ 
2003, \apj, 591, 936 
  
\bibitem[Snider et al.(2001)]{2001ApJ...562..528S} Snider, S., Allende 
Prieto, C., von Hippel, T., Beers, T.~C., Sneden, C., Qu, Y., \& Rossi, S.\ 
2001, \apj, 562, 528 

\bibitem[Spite \& Spite(1982)]{1982A&A...115..357S} Spite, F., \& Spite, 
M.\ 1982, \aap, 115, 357 
 

\bibitem[Stein \& Nordlund(1998)]{1998ApJ...499..914S} Stein, R.~F., \& 
Nordlund, \AA.\ 1998, \apj, 499, 914 

\bibitem[Tsuji(2005)]{2005ApJ...621.1033T} Tsuji, T.\ 2005, \apj, 621, 1033 
 


\bibitem[Van Winckel(2003)]{2003ARA&A..41..391V} Van Winckel, H.\ 2003, 
\araa, 41, 391 
 
 
 \bibitem[Van Winckel et al.(1995)]{1995A&A...293L..25V} Van Winckel, H., 
Waelkens, C., \& Waters, L.~B.~F.~M.\ 1995, \aap, 293, L25 

\bibitem[Westin et al.(2000)]{2000ApJ...530..783W} Westin, J., Sneden, C., 
Gustafsson, B., \& Cowan, J.~J.\ 2000, \apj, 530, 783 

\bibitem[Willemsen et al.(2005)]{2005A&A...436..379W} Willemsen, P.~G., 
Hilker, M., Kayser, A., \& Bailer-Jones, C.~A.~L.\ 2005, \aap, 436, 379 
 
  
\bibitem[Young \& Arnett(2005)]{2005ApJ...618..908Y} Young, P.~A., \& 
Arnett, D.\ 2005, \apj, 618, 908 
 
\end{thebibliography}
\end{document}